\documentclass[11pt]{article}
\usepackage[sfdefault]{carlito} 
\usepackage{geometry}
\usepackage{setspace}
\usepackage{titlesec}
\usepackage{hyperref}
\usepackage{graphicx}
\usepackage{amsmath}
\usepackage{bm}
\usepackage{wrapfig}
\usepackage{xcolor}
\usepackage{authblk}
\usepackage[font=small,labelfont=bf]{caption}

\usepackage{amssymb,bbm,braket,xspace}
\newcommand{\mat}[1]{\mathbf{\underline{#1}}}
\newcommand{\CODE}[1]{\texttt{#1}\xspace}
\newcommand{\FHIaims}{\CODE{FHI-aims}}

\newcommand{\controlin}{\CODE{control.in}}

\renewcommand{\vec}[1]{\mathbf{#1}}
\renewcommand{\t}[1]{\text{#1}}

\newcommand{\ChapFundamentals}[0]{1.1}

\newcommand{\ChapDFPT}[0]{5.1}

\titleformat{\section}{\normalfont\fontsize{11}{13}\bfseries\sffamily}{\thesection}{1em}{}
\titleformat{\subsection}{\normalfont\fontsize{11}{13}\bfseries\sffamily}{\thesubsection}{1em}{}
\titleformat{\title}{\normalfont\fontsize{14}{16}\bfseries\sffamily}{}{0em}{}

\title{Polarisation, Born Effective Charges, and Topological Invariants via a Berry-Phase Approach}
\author[1,2]{Christian Carbogno \thanks{carbogno@fhi-berlin.mpg.de}}
\author[1,3]{Nikita Rybin}
\author[1]{Sara Panahian Jand}
\author[1,4]{Alaa Akkoush}
\author[1,5,6]{Carlos Mera Acosta}
\author[1]{Zhenkun Yuan}
\author[4]{Mariana Rossi}
\affil[1]{The NOMAD Laboratory at the Fritz-Haber-Institut der Max-Planck-Gesellschaft Faradayweg 4-6, 14195, Berlin, Germany}
\affil[2]{{\it Current address:} Theory Department, Fritz Haber Institute of the Max Planck Society, Faradayweg 4-6, 14195 Berlin, Germany}
\affil[3]{{\it Current address:} Skolkovo Institute of Science and Technology, Bolshoi bulvar 30, build.1, 121205, Moscow, Russia}
\affil[4]{Max Planck Institute for the Structure and Dynamics of Matter, 22761 Hamburg, Germany}
\affil[5]{Center for Natural and Human Sciences, Federal University of ABC, Santo André, SP, Brazil}
\affil[6]{{\it Current address:} Faculty of Engineering, University of La Sabana, Chía 250001, Colombia}

\date{}

\begin{document}

\maketitle

\section*{Summary}
The Berry connection~$\vec{A}_{mn}(\vec{k})$, the Berry curvature~$\vec{\nabla}\times\vec{A}_{mn}(\vec{k})$, and the Berry phase~$\varphi=\oint \vec{A}_{mn}(\vec{k})\cdot d\vec{k}$ are key properties describing the reciprocal-space topology, here 
the connection between two electronic states labeled~$m$ and~$n$. They provide a profound link between the phase of a quantum wave function and macroscopic observables as well as material properties. Most prominently, these quantities are central to our understanding of topological materials and provide a route to classify phases in terms of topology~\cite{Bansil.2016}. For instance, all the aforementioned quantities 
enter the definition of the topological-invariant~$\mathbb{Z}_2$ as given by Fu and Kane~\cite{Fu.2006}:
\begin{equation}
\label{EQ_Z2}
\mathbb{Z}_2=\frac{1}{2 \pi} \sum_{n}^\t{occ}\left\{\,\oint\limits_{\partial B} \vec{A}_{nn}(\vec{k})\cdot d\vec{k} - \int\limits_B \left[\vec{\nabla}\times\vec{A}_{nn}(\vec{k})\right] \; d^2\vec{k} \,\right\} \;\t{ mod } 2
\end{equation}
Here, $B$ is half the Brillouin zone~(BZ) and $\partial B$ its boundary, the sum runs over all occupied states, and the gauge of~$\vec{A}_{nm}(\vec{k})$ 
is constrained to respect time-reversal symmetry. 

Furthermore, these quantities play a fundamental role for computing the polarization in periodic systems~\cite{Resta.1992,King-Smith.1993}, or, more precisely, for calculating the polarization density~$\vec{P}$ and its derivatives, the Born effective charges~$\mat{Z}_{I}^*$, via 
\begin{equation}
\label{EQ_pol}
\vec{P} = \frac{1}{V} \left[ \sum_I Z_I \mathbf{R}_I - \frac{eV}{(2 \pi)^3} \sum_n^\t{occ}\oint\limits_\t{BZ} \vec{A}_{nn}(\vec{k}) \; d^3\mathbf{k} \right] \;\; \text{mod } \vec{P}_0 \quad\t{and}\quad Z_{\alpha,I\beta}^*=\frac{V}{e}\frac{\partial P_\alpha}{\partial {R}_{I\beta}} \;.
\end{equation}
Here, $V$ denotes the volume of the unit cell, $\vec{P}_0=\frac{e}{V}(|\vec{a}_1|,|\vec{a}_2|,|\vec{a}_3|)$ the polarization quanta along the lattice vectors~$\vec{a}_\alpha$, $Z_I$ the nuclear charge, $\vec{R}_I$ the nuclear positions, and $e$~the elementary charge. The first term describes the trivial contribution of the bare nuclei, whereas the second term covers the contributions stemming from the electronic states~$\Psi_n(\vec{k})$. 
Besides providing the theoretical foundation for understanding the quantization of adiabatic charge transport~\cite{Thouless.1983}, the polarization 
is a key property for describing the electrodynamics in solids,~e.g.,~for modeling light-matter interactions and for studying ferroelectric and piezoelectric effects.

In addition, these quantities play a central role in the assessment of currents, fluxes, magnetization, and, last but not least, in the transformation of delocalized electronic wave functions into a localized Wannier basis~\cite{Blount.1962,Marzari.1997,Marzari.2012}. We refer the interested reader to Ref.~\cite{Vanderbilt.2018} for a more throughout discussion of all these effects.

\section*{Current Status of the Implementation}
To compute all the aforementioned material properties, the fundamental quantity that needs to be calculated in a first-principles code 
is the Berry connection~\cite{Blount.1962}:
\begin{eqnarray}\label{eq:berry_connection}
A_{mn}(\vec{k}) = i \braket{u_m(\vec{k})| \partial u_n(\vec{k})/\partial \vec{k} } \;.
\end{eqnarray}
Here, $u_l(\vec{k})= \exp(-i\vec{k}\vec{r})\Psi_{l}(\vec{k},\vec{r})$ is the lattice-periodic part of the electronic wave function~$\Psi_{l}(\vec{k},\vec{r})$ for state~$l$ with wave 
vector~$\vec{k}$ and $\braket{\cdot|\cdot}$ denotes the scalar product in Hilbert space. Before evaluating this definition,
let us remind that \FHIaims uses a Bloch-like representation of the wave functions
\begin{eqnarray}
\Psi_{n}(\vec{k},\vec{r}) = \sum_{\mu} C_{\mu}(n,\vec{k}) \chi_{\mu}(\vec{k},\vec{r}) \quad \text{with} \quad
\chi_{\mu}(\vec{k},\vec{r}) = \sum_{N} \exp(i \vec{k}\vec{L}_N) \phi_{\mu,N}(\vec{r}-\vec{R}_\mu - \vec{L}_N) \;,
\end{eqnarray}
as detailed in Contrib.~\ChapFundamentals. Here, $C_{\mu}(n,\vec{k})$ are the Kohn-Sham expansion 
coefficients and $\phi_{\mu,N}(\vec{r}-\vec{R}_\mu - \vec{L}_N)$ are the numeric atomic orbitals~(NAOs)
associated to the basis function with index $\mu$ for the periodic image in the cell~$\vec{L}_N$ of the 
atom located at~$\vec{R}_\mu$. With that, the Berry connection can be expressed as
\begin{eqnarray}
\label{EQ_connection}
A_{mn}(\vec{k}) & = &
   \overbrace{i \sum_{\nu}\sum_{\mu}  C_{\nu,n}^*(\vec{k}) \frac{\partial C_{\mu,n}(\vec{k})}{\partial \vec{k}} S_{\mu\nu}(\vec{k})}^{A_{mn}^{(1)}(\vec{k})} \\
&& - \underbrace{\sum_{\nu}\sum_{\mu}  C_{\nu,n}^*(\vec{k})C_{\mu,n}(\vec{k}) \left[ {D}_{\mu,\nu}(\vec{k}) - \vec{R}_\mu  S_{\mu\nu}(\vec{k}) \right]}_{A_{mn}^{(2)}(\vec{k})} \;.\nonumber
\end{eqnarray}
The first term in Eq.~(\ref{EQ_connection}),~i.e.,~$\vec{A}_{mn}^{(1)}(\vec{k})$, denotes the gauge-dependent Berry-connection term, which here includes the overlap matrix~$S_{\mu\nu}(\vec{k}) = \sum_{N} e^{i \vec{k} \vec{L}_N}  \braket{\phi_{\nu,0}| \phi_{\mu,N}}$ due to the non-orthogonality of the employed NAO basis set. For evaluating the associated contribution to the Berry phase,~i.e.,~the closed-path integrals required for Eq.~(\ref{EQ_Z2}) and Eq.~(\ref{EQ_pol}), the reciprocal-space path is discretized on $K$ points~$(\vec{k}_1,\vec{k}_2,\cdots,\vec{k}_K)$, whereby the initial and final point are equivalent with respect to the BZ's periodicity~$\vec{k}_1=\vec{k}_K \text{ mod } \frac{2\pi}{V}$. By 
expressing the $\vec{k}$-derivatives as two-point finite-differences, one obtains~\cite{Marzari.1997} 
\begin{equation}
\label{EQ_disc}
\sum_n^\t{occ}\oint\limits_\t{BZ} \vec{A}_{nn}^{(1)}(\vec{k}) \cdot d\vec{k} = - \text{Im}\;\ln\left[ \text{det}\left(\mat{M}_{0,1} \cdot \mat{M}_{1,2} \cdots \mat{M}_{K-2,K-1} \cdot \mat{M}_{K-1,0}\right) \right] \;.
\end{equation}
Note that the Kohn-Sham coefficients~$\mat{C}^\t{occ}(\vec{k})$ used for computing the matrices
$$\mat{M}_{a,b} =  \mat{C}^{\t{occ}^\dag}(\vec{k}_a)  \mat{S}(\vec{k}_a) \mat{C}^\t{occ}(\vec{k}_b)$$ 
entering the above expression only cover the subspace of occupied states.

The second term entering Eq.~(\ref{EQ_connection}),~i.e.,~$\vec{A}_{mn}^{(2)}(\vec{k})$, is gauge-invariant and features the matrix
\begin{eqnarray}
D_{\mu\nu}(\vec{k}) & = & - \sum_{N} e^{i \vec{k} \vec{L}_N}  \braket{\phi_{\nu,0}|[\vec{r}-\vec{R}_\mu-\vec{L}_N]| \phi_{\mu,N}} \;,
\end{eqnarray}
which captures the contributions of the NAO basis functions. Since this form exhibits the exact same periodicity as the overlap matrix~$\mat{S}(\vec{k})$, it can be integrated up using the real-space routines already present in \FHIaims~\cite{Havu.2009,Knuth.2015}. Similarly, 
the associated contribution to the Berry-phase required for Eq.~(\ref{EQ_Z2}) and Eq.~(\ref{EQ_pol}) can be computed straightforwardly 
by performing the trace over occupied states and by numerically integrating~$A_{mn}^{(2)}(\vec{k})$ along the exact same path used in Eq.~(\ref{EQ_disc}).

\begin{wrapfigure}{r}{0.55\textwidth}
  \centering
    \includegraphics[width=0.53\textwidth]{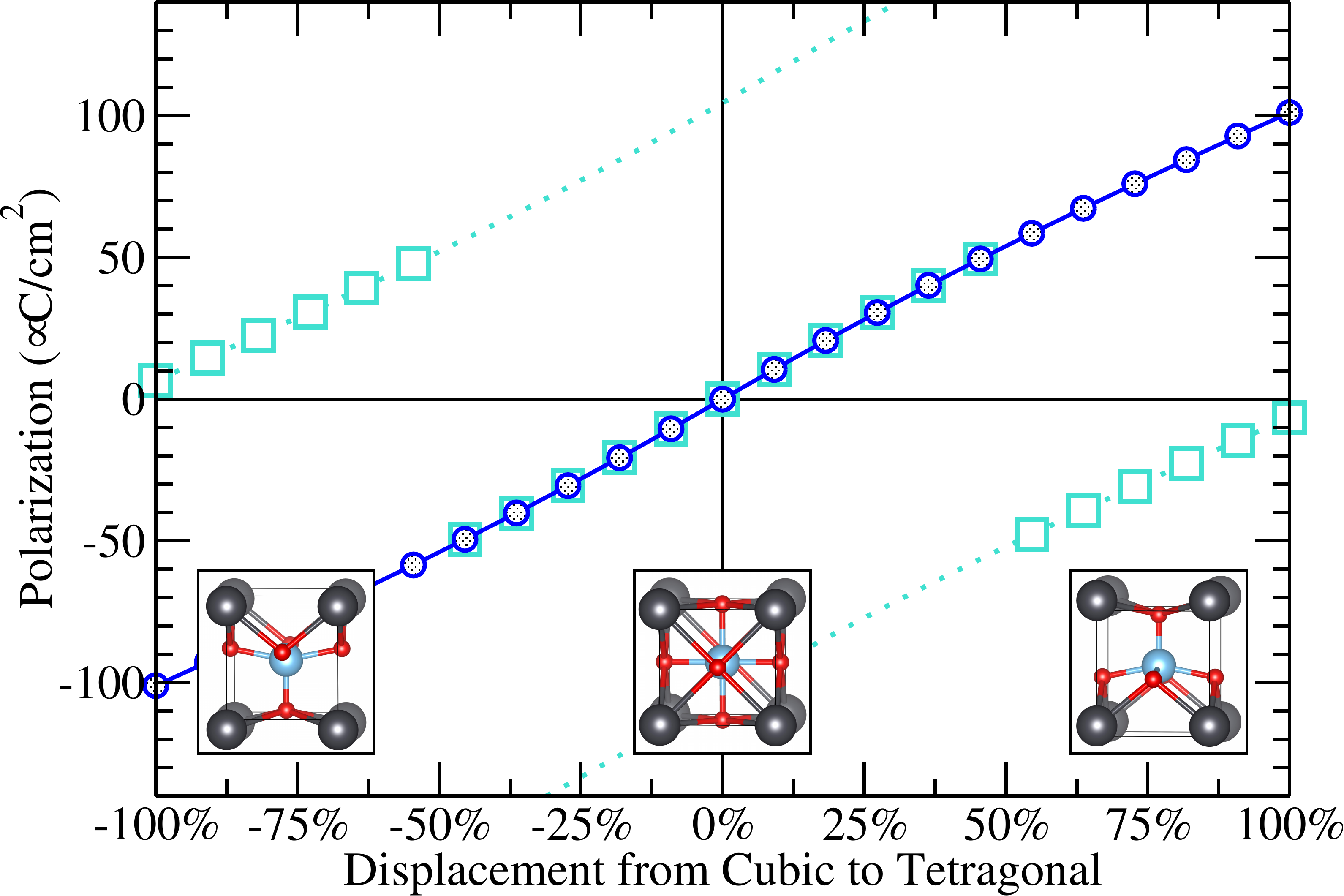}
    \caption{Polarisation~($24\times 8 \times 8$ $\vec{k}$-points for the Berry-phase) of PbTiO$_3$~(HSE06, 8$^3$  $\vec{k}$-points) 
    along the minimum-energy path connecting the tetragonal, symmetry-degenerate~$P4mm$ and the cubic, centrosymmetric~$Pm\bar{3}m$ structure.
    Cyan squares denote the bare output, blue circles the ``branch-matched'' polarization, for which the discontinuities associated to the $\text{mod } \vec{P}_0$-operation
    are resolved.}
    \label{FIG_PBTIO3}
\end{wrapfigure}
In the current implementation, the polarization is calculated in the basis of the reciprocal-lattice vectors. To perform the $\vec{k}$-derivative 
along the reciprocal-lattice vector of interest, the closed-loop path is chosen parallel to it. The remaining integrations~$\int d^2\vec{k}$ perpendicular
to this path are performed by splining the Berry phases. As an example for such a calculation, Fig.~\ref{FIG_PBTIO3} shows 
the polarization of PbTiO$_3$, here for smoothly interpolated geometries and lattices between the tetragonal, symmetry-degenerate~$P4mm$ equilibrium structures 
and the cubic, centrosymmetric~$Pm\bar{3}m$ structure. While the latter structure must have a vanishing polarization due to symmetry, the tetragonal configurations 
do not. This highlights that the absolute values of the polarization are meaningless, only relative differences of the polarization matter. To evaluate such differences,
``branch-matching''~\cite{Spaldin.2012},~i.e.,~ensuring that the actual polarization values lay on the branch associated to the same multiples of~$\vec{P}_0$ value,
is crucial\footnote{Note that an internal branch matching in \FHIaims is already performed for splining and integrating over the perpendicular $\vec{k}$-directions.}.

\begin{wrapfigure}{l}{0.55\textwidth}
  \centering
    \includegraphics[width=0.53\textwidth]{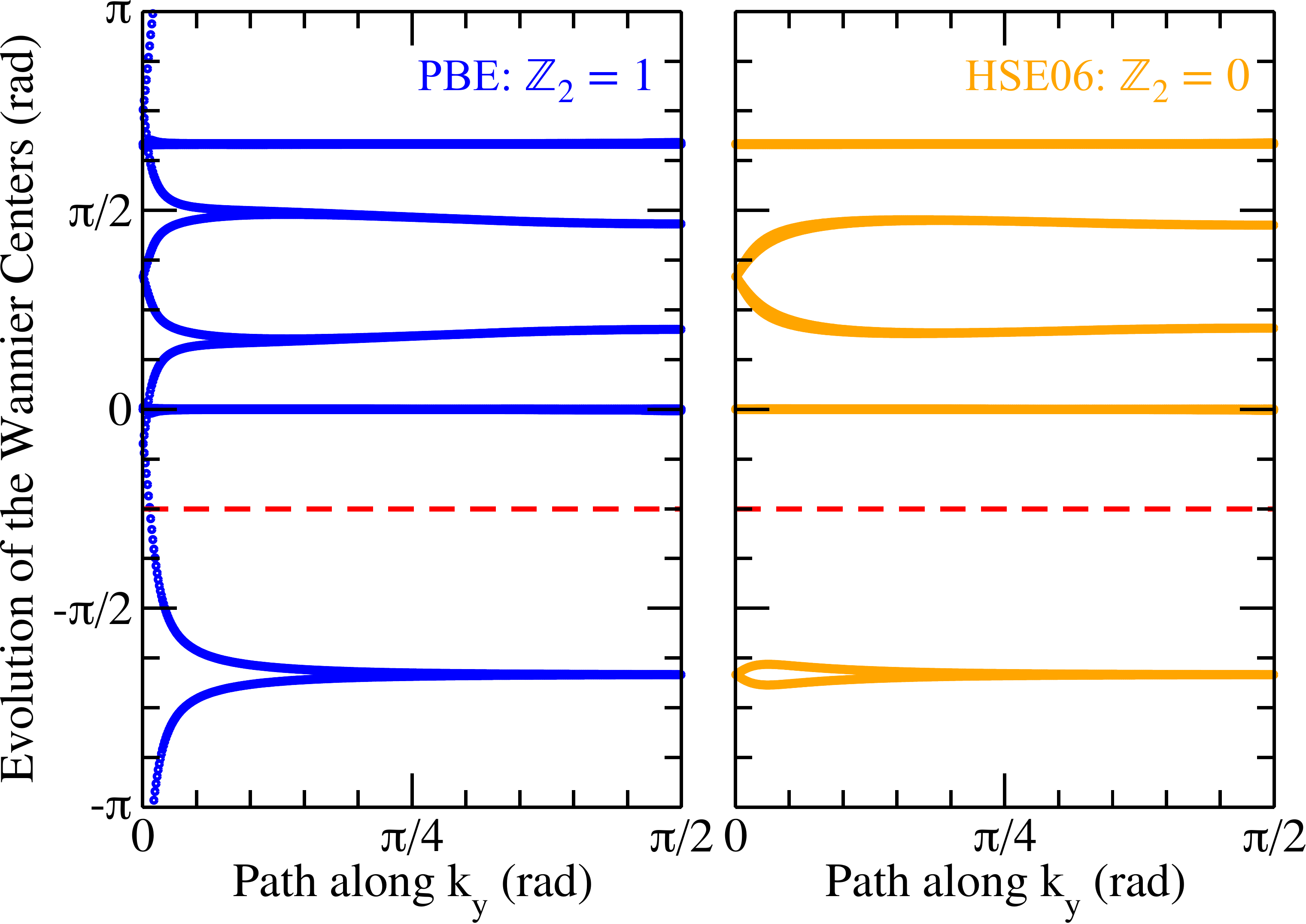}
    \caption{Evolution of the Wannier Centers of Charge for the functionalized 2D-honeycomb structure of GeF$_2$, as computed with PBE~(left) and HSE06~(right) along $\vec{k}_2=k_y$. Both calculations 
    use spin-orbit coupling~\cite{Huhn.2017}. As indicated by     the red, dashed line, a continuous path along $k_y$ must cross the Wanniers Centers an odd~(PBE) viz.~even~(HSE06) 
    number of times, resulting in $\mathbb{Z}_2=1$ and $\mathbb{Z}_2=0$, respectively. This showcases the influence of the exchange-correlation functional on topological invariants~\cite{Si.2014,Matthes.2016}.}
    \label{FIG_WCC}
\end{wrapfigure}

For the evaluation of the $\mathbb{Z}_2$ invariant, the current implementation follows the formalism proposed in Refs.~\cite{Soluyanov.2011,Yu.2011}, which is equivalent to the
definition given in Eq.~(\ref{EQ_Z2}), but does not require gauge-fixing. In practice, it requires to track the evolution of the individual Wannier centers
\begin{equation}
\label{EQ_WCC}
\vec{X}_n(\vec{k}_2) =  \oint\limits_{-\pi}^{\pi} A_{nn}(\vec{k}_1) \cdot d\vec{k}_1 \quad\text{mod}\; 2\pi
\end{equation}
across a path in the BZ described by~$\vec{k}_2$. In practice, one evaluates the line-path integral in Eq.~(\ref{EQ_WCC}) for varying values of~$\vec{k}_2$. Each of these 
integrals is solved for all occupied states as discussed above,~i.e.,~by using $A_{mn}^{(1)}(\vec{k})$ as given by Eq.~(\ref{EQ_disc}) and~$A_{mn}^{(2)}(\vec{k})$ along 
a discretized path~$\vec{k}_1 \perp \vec{k}_2$. The determinant viz. trace is, however, not evaluated. Rather, the obtained matrix is diagonalized 
and the complex phase of the resulting eigenvalues is then tracked, as the example in Fig.~\ref{FIG_WCC} shows. 
If an arbitrary continuous line across the whole $\vec{k}_2$-axis crosses the evolution of the Wannier centers an even number of times~$\mathbb{Z}_2$ is 0 and otherwise~1.
Similarly, this can be judged by tracking the largest gap between the individual Wannier centers~\cite{Soluyanov.2011}.

\section*{Usability and Tutorials} 
For the evaluation of the polarization, it is sufficient to add one keyword to the \controlin file:
\begin{center}
\texttt{output polarization $\alpha$ $n_1$ $n_2$ $n_3$}
\end{center}
For instance, the line \texttt{output polarization 2 5 10 5} will compute the polarization along the reciprocal-lattice vector number~$2$ using a
grid of~$5\times10\times5$ $\vec{k}$-points in the BZ. As the example highlights, the discretization used along the reciprocal-lattice vector of
interest,~i.e.,~the one along which the closed-path integral in Eq.~(\ref{EQ_disc}) is performed, usually requires denser grids. For convenience,
multiple \texttt{output polarization} statements can be combined and, if all three directions are requested, the code will also report the 
polarization in Cartesian coordinates. For the evaluation of Born effective charges~$\mat{Z}_I$, a Python script~\CODE{BEC.py} is provided 
in the utilities folder of the \FHIaims distribution to perform the required derivatives, cf.~(\ref{EQ_pol}) via finite differences. Although these functionalities are
rather self-explanatory, a tutorial is provided at \texttt{https://fhi-aims-club.gitlab.io/tutorials/phonons-with-fhi-vibes/}, which also showcases
how the computed Born effective charges can be used to account for long-range dipole interactions in the calculations of phonon spectra in
polar crystals, see~\cite{Baroni.2001} and references therein.

Similarly, the evaluation of $\mathbb{Z}_2$ viz.~of the evolution of the Wannier centers of charge only requires to add one keyword to the \controlin file:
\begin{center}
\texttt{output Z2\_invariant $\gamma$ $n_\parallel$ $n_\perp$}
\end{center}
Here, $\gamma$ is an index that encodes which Cartesian planes shall actually be targeted,~e.g,~$\gamma=1$ implies that Eq.~(\ref{EQ_WCC}) is evaluated along
the first reciprocal-lattice vector using a $\vec{k}$-discretization of $n_\parallel$ points and that this procedure is repeated for $n_\perp$ 
closed-paths that have equidistant~$k_2 \in [0,\pi]$ and~$k_3=0$. The latter scan over $k_2$ can, for instance, be used to discern strong from weak topological
insulators~\cite{Fu.2007}.

Eventually, let us note that the implementation supports all exchange-correlation functionals, i.e.,~all semi-local and hybrid functionals, as well as
spin-orbit coupling as described in Ref.~\cite{Huhn.2017}. Also, given that this functionality targets unit cells and requires rather dense $\vec{k}$-grids
the parallelization occurs over $\vec{k}$-points using \CODE{LAPACK} for the compute-intense linear-algebra operations.

\section*{Future Plans and Challenges}
So far, the existing implementation has proven useful, accurate, and performant for targeting relatively simple materials with few~($<100$) atoms in the unit 
cell.  However, there is increased scientific and technological interest in targeting materials with 
structural or compositional disorder,~e.g.,~for alloyed topological insulators featuring thousands of atoms in the unit cell~\cite{Cao.2020}. For such kind 
of systems, the current $\vec{k}$-point-based parallelization strategy  is not efficient. Rather, support for distributed linear algebra~(ScaLAPACK) is 
needed and is currently being pursued.

Furthermore, the described Berry-phase approach does not only give access to polarization, Born effective charges, and topological invariants, but to a multitude 
of other material properties, as described in the introduction. In this context, a systematic interface between the methodologies described in this contributions
and the density-functional perturbation theory implementation described in Contrib.~\ChapDFPT is desirable for the accelerate assessment of response 
properties,~e.g.,~piezoelectric tensors, but also Born effective charges, or other properties pivotal for electrodynamics viz.~light-matter interactions. 
Another route that is being exploited is the machine-learning of the polarization for systems with a reduced number of atoms using a local representation. These 
simulations allow us to treat much larger unit cells. While learning the polarization requires care because of dealing with a topological quantity~\cite{Xie.2022}, our goal is 
to target models that can be used in the context of nuclear dynamics with light-matter coupling, as reported in Ref.~\cite{Litman.2024}. 
Finally, let us emphasize that the implemented infrastructure also makes a transformation to Wannier functions straightforward. While
this functionality would not be particularly useful with \FHIaims itself, given that the NAOs already provide a localized representation, it would be beneficial
for interfacing to other community codes based on a Wannier representation. For instance, this would give access to all the functionality provided 
by,~e.g.,~\CODE{Wannier90}~\cite{Pizzi.2020}, \CODE{EPW}~\cite{Ponce.2016}, and \CODE{Perturbo}~\cite{Zhou.2021}, and, in turn, enable more systematic, community-wide
benchmarks and collaborations across ``code-boundaries''.

\end{document}